# Spin-orbit torque in Cr/CoFeAl/MgO and Ru/CoFeAl/MgO epitaxial magnetic heterostructures

Zhenchao Wen, Junyeon Kim, Hiroaki Sukegawa, Masamitsu Hayashi, and Seiji Mitani [a]

*National Institute for Materials Science (NIMS), 1-2-1 Sengen, Tsukuba 305-0047, Japan*

We study the spin-orbit torque (SOT) effective fields in Cr/CoFeAl/MgO and Ru/CoFeAl/MgO magnetic heterostructures using the adiabatic harmonic Hall measurement. High-quality perpendicular-magnetic-anisotropy CoFeAl layers were grown on Cr and Ru layers. The magnitudes of the SOT effective fields were found to significantly depend on the underlayer material (Cr or Ru) as well as their thicknesses. The damping-like longitudinal effective field ($\Delta H_L$) increases with increasing underlayer thickness for all heterostructures. In contrast, the field-like transverse effective field ($\Delta H_T$) increases with increasing Ru thickness while it is almost constant or slightly decreases with increasing Cr thickness. The sign of $\Delta H_L$ observed in the Cr-underlayer devices is opposite from that in the Ru-underlayer devices while $\Delta H_T$ shows the same sign with a small magnitude. The opposite directions of $\Delta H_L$ indicate that the signs of spin Hall angle in Cr and Ru are opposite, which are in good agreement with theoretical predictions. These results show sizable contribution from SOT even for elements with small spin orbit coupling such as 3$d$ Cr and 4$d$ Ru.

**I. INTRODUCTION**

Spin-orbit torque (SOT) has drawn much interest for achieving electrical manipulation of magnetization available for high-performance spintronic devices, such as magnetic tunnel junctions (MTJs),[1-3] magnetic nano-oscillators,[4] and chiral domain-wall devices[5-9]. In nonmagnetic metal/ferromagnetic metal (NM/FM) heterostructures, spin Hall effect (SHE)[10,11] and/or Rashba-Edelstein effect[12] produces nonequilibrium spin accumulation at the interface, which exerts SOT on the magnetization of a ferromagnetic layer.[13,14] The SOTs acting on the magnetization of an FM layer can be considered as a form of an effective field, along (longitudinal field) or transverse (transverse field) to the direction of current flow when the FM layer is perpendicularly magnetized.[15] Manipulation of magnetization by SOT can be efficient and provide a unique way to develop low-consumption and high-performance spintronic devices.

---

[a] Electronic mail: Seiji.Mitani@nims.go.jp

To date, the SOTs have been intensively studied in various NM/FM bilayers, where the NMs are usually 5$d$ heavy metals and/or their alloys, such as Ta,[1,15-17] Pt,[7,18-20] W,[21,22] Hf,[23] Ir-doped Cu,[2] and Bi$_2$Se$_3$,[24] and FMs are typically CoFeB, Co, or NiFe. SOTs typically arise in heterostructures that contain heavy metals with strong spin orbit coupling and in which the structural inversion symmetry is broken. Interestingly, the longitudinal field, which originates from the anti-damping component of the spin-Hall spin torques, shows opposite directions between Ta/CoFeB/MgO and Pt/Co/Oxide heterostructures because of the opposite signs of spin Hall angles between Pt and Ta underlayers.[7] The relationship between the direction of longitudinal field and the sign of spin Hall angle was also supported in a W/CoFeB/MgO heterostructure.[21,22] Theoretically, spin Hall angles of 4$d$ and 5$d$ transition metals were investigated by tight-binding model calculations,[25,26] which predict the sign and magnitude of spin Hall angles systemically depend on the number of $d$-orbital electrons. More recently, SOTs arising from SHE in the antiferromagnetic metals (AFM) were exploited,[27,28] which promises the external-field-free switching in AFM/FM based spintronic devices.[29] Studies of SOTs in relatively light elements are rarely performed so far due to the expected weak spin-orbit interaction.

In this work, we studied the SOT in Cr and Ru based magnetic heterostructures using the adiabatic harmonic Hall measurement[15]. Perpendicular magnetic anisotropy (PMA) CoFeAl (CFA) layers were grown on Cr and Ru layers[30,31] for examining SOTs. The utilization of epitaxial CFA eliminates boron diffusion into the underlayers in order to provide a clean system to study. The magnitudes of the SOT effective fields were found to significantly depend on the Cr and Ru elements as well as their thicknesses. The sign of the damping-like longitudinal effective field ($\Delta H_L$) observed in the Cr-underlayer devices is opposite from that in the Ru-underlayer devices while the field-like transverse effective field ($\Delta H_T$) shows the same sign with a small magnitude. The opposite directions of $\Delta H_L$ indicate that the signs of spin Hall angle in Cr and Ru are opposite, which are in good agreement with theoretical predictions[25].

## II. EXPERIMENTAL

The multilayer stacks with the structure of Ru(0−10)/CFA(1)/MgO(2)/Ta(1) (sample A), Cr(0−10)/CFA(1)/MgO(2)/Ta(1) (sample B), and Cr(0.5)/CFA(0.8)/MgO(10) (sample C) (unit: nm) were deposited on MgO (001) substrates at room temperature (RT) by an ultrahigh vacuum magnetron sputtering system with a base pressure of around $4 \times 10^{-7}$ Pa. The stacks were annealed at 325 °C for 1 hour in a vacuum furnace after deposition in order to achieve PMA of the CFA thin films. The magnetization hysteresis ($M$-$H$) loops under in-plane and out-of-plane magnetic fields were measured at RT using a vibrating sample magnetometer (VSM). After characterizing the magnetic properties of the CFA thin films, Hall bar structures were micro-fabricated by conventional UV lithography combining with Ar ion milling, and contact electrodes with Ta (10 nm)/Au (100 nm) bilayers were made by a lift-off process after formation of the Hall bars. The current-induced effective fields were investigated by an adiabatic harmonic measurement.[15,19,32]

## III. RESULTS AND DISCUSSION

The in-plane and out-of-plane *M-H* loops of the epitaxial heterostructures of sample A and sample B are shown in Figs.1 (a) and (b), respectively. The perpendicular magnetization of CFA thin films is achieved on the Ru or Cr underlayer with MgO capping layers. Since bulk CFA has a low magnetocrystalline anisotropy due to the cubic crystalline structure, the PMA achieved in the thin CFA films indicates that the interface-induced perpendicular anisotropy plays a significant role, which can be explained by the hybridization between Fe-3$d$ and O-2$p$ electron orbitals at the CFA/MgO interface.[33] The PMA energy density ($K_u$) of $3.0 \times 10^6$ erg/cm$^3$ is obtained in sample A, whereas, a relative smaller $K_u$ of $1 \times 10^6$ erg/cm$^3$ is observed in sample B. The $K_u$ was calculated from the area enclosed between the out-of-plane and in-plane magnetic hysteresis loops. The enhanced PMA of the CFA thin film on the Ru underlayer may be attributed to the improved CFA/MgO interfacial structure owing to the novel 4-fold-symmetry Ru structure.[31]

After characterizing the magnetic properties of the CFA thin films, Hall bar structures were micro-fabricated for investigating SOTs induced effective fields. The transverse ($\Delta H_T$) and longitudinal ($\Delta H_L$) effective fields are obtained by the following equations in which the planar Hall effect (PHE) contribution is included.[34]

$$\Delta H_{T(L)} = \left( \Delta H'_{T(L)} \pm 2\xi \Delta H'_{L(T)} \right) / \left( 1 - 4\xi^2 \right), \quad (1)$$

$$\Delta H'_{T(L)} = -2 \left( \frac{\partial V_{2\omega}}{\partial H_{T(L)}} \right) \Big/ \left( \frac{\partial^2 V_\omega}{\partial H_{T(L)}^2} \right),$$

$$\xi = \Delta R_P / \Delta R_A.$$

where $V_\omega$ ($V_{2\omega}$) is the first (second) harmonic signal, $H_{T(L)}$ is an applied in-plane field directed transverse (parallel) to the current flow, $\Delta R_P$ and $\Delta R_A$ are Hall resistances due to PHE and AHE, respectively. The illustrations of Hall bar devices with the measurement set-up are shown in Figs.2 (a) and (b). For all samples, the first harmonic signal $V_\omega$ shows a typically parabolic feature as a function of external field, whereas, the second harmonic signal $V_{2\omega}$ exhibits a linear response to the external fields. Figures 2(a) and (b) show representative feature of $H_{T(L)}$ dependence of $V_{2\omega}$ measured in sample B (Cr thickness is 8.0 nm and the ac voltage $V_{IN}$ is 7.5 V), where $+M_z$ and $-M_z$ indicate the orientation of out-of-plane magnetization of CFA thin films. Under the transverse external field, the sign of the slopes of $V_{2\omega}(H_T)$ reverses with the magnetization orientations, whereas their sign is the same for the longitudinal external field. Figures 2 (c) and (d) show the transverse and longitudinal effective field $\Delta H'_{T(L)}$ before the PHE correction as a function of the ac voltage $V_{IN}$, which was measured in sample B (Cr thickness: 8.0 nm). $\Delta H'_T$ is independent of the magnetization direction while $\Delta H'_L$ reverses. The magnitude of $\Delta H'_{T(L)}$ linearly increases with the increase of input voltage. In addition, the sign of $\Delta H'_L$ observed in Cr-underlayer devices (samples B and C) is opposite from that in Ru-underlayer devices (sample A) where $\Delta H'_T$ shows the same sign with a small magnitude.

Since the measured Hall voltage simultaneously includes the contributions from both AHE and PHE, the Hall resistance $R_H$ can be expressed by,

$$R_H = \frac{1}{2}\Delta R_A \cos\theta + \frac{1}{2}\Delta R_P \sin^2\theta \sin 2\varphi \qquad (2)$$

where we introduce spherical coordinate system as commonly used in physics, $\theta$ is polar angle, $\varphi$ azimuthal angle.[34] We investigated $\Delta R_A$ and $\Delta R_P$ by using physical properties measurement system (PPMS) to correct the effective field. During the measurements, we rotated the sample for a polar angle $\theta$ or an azimuthal angle $\varphi$ with a constant external field of 2 T. Figures 3 (a) and (b) show the polar angle $\theta$ ($\varphi = 0°$) dependence of the AHE resistance $R_H(\theta; \varphi = 0°)$ for sample A and sample B with 8-nm-thick underlayers. A cosine feature is observed, which is consistent with the equation (2). The PHE resistance $R_H(\theta = 90°; \varphi)$ was measured as a function of azimuthal angle $\varphi$ ($\theta = 90°$) in the same samples, as shown in Figs. 3 (c) and (d). A magnified $R_H(\varphi)$ is also shown in the inset of the figures. The period of the PHE resistance is two times larger than that of the AHE resistance, in accordance with equation (2). Thus, $\Delta R_A$ and $\Delta R_P$ are obtained by fitting equation (2). The fitted curves are shown by the solid lines in the corresponding figures. As a result, the amplitude of $\Delta R_P$ is significantly smaller than that of $\Delta R_A$. Figures 3 (e) and (f) show the underlayer thickness dependence of $\xi = \Delta R_P/\Delta R_A$ for samples A and B, respectively. The ratios of $\xi \sim 5\%$ for the Ru underlayer and $\xi \sim 2.5\%$ for the Cr underlayer were observed. Considering about the relationship between the PHE correction and effective fields, the small $\xi$ indicates that the influence of PHE to the effective field is less significant in these material systems.

In order to investigate the effective field quantitatively, $\Delta H_T$ and $\Delta H_L$ are normalized by an applied current density $J$ of $1\times 10^8$ A/cm².[20] Figures 4 (a) and (b) show the underlayer thicknesses dependence of $\Delta H_T/J$ after the PHE correction for samples A and B, respectively. Furthermore, in order to obtain the intrinsic effective field, we also considered the influence of Oersted field generated from the NM layer[34] on the measured transverse effective field $\Delta H_T$. The estimated Oersted fields as a function of underlayer thickness are shown by the solid lines in Figs.4 (a) and (b) for sample A and sample B, respectively. Note that we assume the current flows to the underlayer with a current density of $1\times 10^8$ A/cm². In contract to the devices of Ta/CoFeB/MgO,[15] the measured $\Delta H_T$ in the present samples is comparable or smaller than the estimated Oersted field.

The effective field from the SOTs is obtained by correcting the PHE contribution and subtracting the Oersted field. Figures 5 (a-d) show the underlayer thickness dependence of the corrected transverse and longitudinal effective fields. The results show that the magnitude of effective fields depends on the underlayer thickness. The magnitude of $\Delta H_L/J$ increases with increasing underlayer thickness for all samples, whereas $\Delta H_T/J$ exhibits a different underlayer thickness dependence for the Ru- and Cr-devices. The magnitude of $\Delta H_T/J$ in sample A increases with increasing Ru underlayer thickness; however, it is almost constant or slightly decreases with increasing Cr underlayer thickness in sample B and C. The thickness dependence of the longitudinal effective field $\Delta H_L/J$ in the Ru- and Cr-devices indicates that its origin is predominantly the SHE via the anti-damping torque. On the other hand, the origin of the transverse effect field $\Delta H_T/J$ is not clear. The underlayer thickness dependent $\Delta H_T/J$ may be caused by the spin-Hall spin torque, and the thickness independent $\Delta H_T/J$ for the devices with thin underlayer could be attributed to an interfacial effect, e.g. Rashba effect. The spin-Hall spin torque and the interfacial torque for $\Delta H_T/J$ in sample A seem to be comparable in size whereas the spin-Hall spin torque for $\Delta H_T/J$ in sample B seems to be nearly zero. Furthermore, we find that the directions of $\Delta H_L/J$ are opposite between the Ru- and Cr-underlayer devices. Both longitudinal and transverse components of the effective field in sample A point along the same direction with those reported in Pt-underlayered devices[7,35] whereas the direction of $\Delta H_L/J$ from sample B and C is the same with that in Ta-, Hf-, and W-underlayer devices.[9] The direction of the damping-like (longitudinal) effective field indicates the sign of spin Hall angle of Ru (Cr) is the same with that of Pt (Ta, Hf and W), which corresponds to a positive (negative) spin Hall angle. This is in agreement with theoretical calculations.[25,26] The obtained effective fields have a relatively small value by comparing with the reported values of Ta/CoFeB/MgO,[15] which clearly indicates the NM/FM interface plays an important role for SOTs even though the FM/MgO interface defines the PMA.

Finally, we quantitatively evaluate the spin Hall angle ($\theta_{SHE}$) in both Cr and Ru layers using the following equation,[35]

$$\theta_{SHE} = \frac{J_S}{J_{C\_NM}} = \frac{2eM_s t_{FM}}{\hbar} \frac{\Delta H_L}{J_{C\_NM}} \qquad (3)$$

where $J_s$ and $J_{c\_NM}$ represent the spin and change currents in the buffer layer; $e$ and $\hbar$ are the electron charge and the reduced Planck constant; $M_s$ and $t_{FM}$ are the saturation magnetization (~1000 emu/cm$^3$) and the thickness of CFA. We assume the charge current mostly flows in the underlayer and spin current is conserved when passing the NM/FM interface. Using the effective field when the underlayer thickness is thick (~8 nm), we estimate the spin Hall angle to be ~0.0056 and ~−0.0070 for Ru and Cr respectively. We note that the obtained values of the spin Hall angle can be underestimated because of the above assumptions.

## IV. CONCLUSIONS

In summary, we studied the current-induced SOTs for perpendicularly magnetized Cr/CFA/MgO and Ru/CFA/MgO epitaxial heterostructures. The SOTs are significantly influenced by the Ru or Cr underlayer as well as the thicknesses. The directions of the anti-damping torque were found to be opposite for the Cr- and Ru-underlayer devices, indicating the sign of the spin Hall angle is opposite for these materials. These results show sizable contribution from the SOT even for elements with small spin orbit coupling such as $3d$ Cr and $4d$ Ru.

## ACKNOWLEDGMENT

This work was supported by JSPS KAKENHI Grant Number 23246006.

## REFERENCES

[1] L. Liu, C.-F. Pai, Y. Li, H. W. Tseng, D. C. Ralph, and R. A. Buhrman, Science **336**, 555 (2012).
[2] M. Yamanouchi, L. Chen, J. Kim, M. Hayashi, H. Sato, S. Fukami, S. Ikeda, F. Matsukura, and H. Ohno, Appl. Phys. Lett. **102**, 212408 (2013).
[3] M. Cubukcu, O. Boulle, M. Drouard, K. Garello, C. Onur Avci, I. Mihai Miron, J. Langer, B. Ocker, P. Gambardella, and G. Gaudin, Appl. Phys. Lett. **104**, 042406 (2014).
[4] L. Liu, C.-F. Pai, D. C. Ralph, and R. A. Buhrman, Phys. Rev. Lett. **109**, 186602 (2012).
[5] A. Fert, V. Cros, and J. Sampaio, Nat. Nanotech. **8**, 152 (2013).
[6] P. P. J. Haazen, E. Murè, J. H. Franken, R. Lavrijsen, H. J. M. Swagten, and B. Koopmans, Nat. Mater. **12**, 299 (2013).
[7] S. Emori, U. Bauer, S.-M. Ahn, E. Martinez, and G. S. D. Beach, Nat. Mater. **12**, 611 (2013).
[8] K.-S. Ryu, L. Thomas, S.-H. Yang, and S. Parkin, Nat. Nanotech. **8**, 527 (2013).
[9] J. Torrejon, J. Kim, J. Sinha, S. Mitani, M. Hayashi, M. Yamanouchi, and H. Ohno, Nat. Commun. **5**, 4655 (2014).
[10] M. I. Dyakonov and V. I. Perel, Phys. Lett. A **35**, 459 (1971).
[11] J. E. Hirsch, Phys. Rev. Lett. **83**, 1834 (1999).
[12] V. M. Edelstein, Solid State Commun. **73**, 233 (1990).
[13] J. C. Slonczewski, J. Magn. Magn. Mater. **159**, L1 (1996).
[14] L. Berger, J. Appl. Phys. **55**, 1954 (1984).
[15] J. Kim, J. Sinha, M. Hayashi, M. Yamanouchi, S. Fukami, T. Suzuki, S. Mitani, and H. Ohno, Nat. Mater. **12**, 240 (2013).
[16] T. Suzuki, S. Fukami, N. Ishiwata, M. Yamanouchi, S. Ikeda, N. Kasai, and H. Ohno, Appl. Phys. Lett. **98**, 142505 (2011).
[17] J. Kim, J. Sinha, S. Mitani, M. Hayashi, S. Takahashi, S. Maekawa, M. Yamanouchi, and H. Ohno, Phys. Rev. B **89**, 174424 (2014).
[18] I. M. Miron, K. Garello, G. Gaudin, P.-J. Zermatten, M. V. Costache, S. Auffret, S. Bandiera, B. Rodmacq, A. Schuhl, and P. Gambardella, Nature **476**, 189 (2011).
[19] U. H. Pi, K. W. Kim, J. Y. Bae, S. C. Lee, Y. J. Cho, K. S. Kim, and S. Seo, Appl. Phys. Lett. **97**, 162507 (2010).
[20] I. M. Miron, G. Gaudin, S. Auffret, B. Rodmacq, A. Schuhl, S. Pizzini, J. Vogel, and P. Gambardella, Nat. Mater. **9**, 230 (2010).
[21] C.-F. Pai, L. Liu, Y. Li, H. W. Tseng, D. C. Ralph, and R. A. Buhrman, Appl. Phys. Lett. **101**, 122404 (2012).
[22] C.-F. Pai, M.-H. Nguyen, C. Belvin, L. H. Vilela-Leão, D. C. Ralph, and R. A. Buhrman, Appl. Phys. Lett. **104,** 082407 (2014).
[23] H. Sukegawa, Y. Miura, S. Muramoto, S. Mitani, T. Niizeki, T. Ohkubo, K. Abe, M. Shirai, K. Inomata, and K. Hono, Phys. Rev. B **86**, 184401 (2012).
[24] A. R. Mellnik, J. S. Lee, A. Richardella, J. L. Grab, P. J. Mintun, M. H. Fischer, A. Vaezi, A. Manchon, E. A. Kim, N. Samarth, and D. C. Ralph, Nature **511**, 449 (2014).
[25] T. Tanaka, H. Kontani, M. Naito, T. Naito, D. S. Hirashima, K. Yamada, and J. Inoue, Phys. Rev. B **77**, 165117 (2008).

[26] H. Kontani, T. Tanaka, D. S. Hirashima, K. Yamada, and J. Inoue, Phys. Rev. Lett. **102**, 016601 (2009).
[27] W. Zhang, M. B. Jungfleisch, W. Jiang, J. E. Pearson, A. Hoffmann, F. Freimuth, and Y. Mokrousov, Phys. Rev. Lett. **113**, 196602 (2014).
[28] D. Qu, S. Y. Huang, and C. L. Chien, Phys. Rev. B **92** 020418(R) (2015).
[29] S. Fukami, C. Zhang, S. DuttaGupta, and H. Ohno, http://arxiv.org/ftp/arxiv/papers/1507/1507.00888.pdf (2015).
[30] Z. Wen, H. Sukegawa, S. Mitani, and K. Inomata, Appl. Phys. Lett. **98**, 242507 (2011).
[31] Z. Wen, H. Sukegawa, T. Furubayashi, J. Koo, K. Inomata, S. Mitani, J. P. Hadorn, T. Ohkubo, and K. Hono, Adv. Mater. **26**, 6483 (2014).
[32] K. Garello, I. M. Miron, C. O. Avci, F. Freimuth, Y. Mokrousov, S. Blugel, S. Auffret, O. Boulle, G. Gaudin, and P. Gambardella, Nat. Nanotech. **8**, 587 (2013).
[33] J. Okabayashi, H. Sukegawa, Z. Wen, K. Inomata, and S. Mitani, Appl. Phys. Lett. **103**, 102402 (2013).
[34] M. Hayashi, J. Kim, M. Yamanouchi, and H. Ohno, Phys. Rev. B **89**, 144425 (2014).
[35] L. Liu, O. J. Lee, T. J. Gudmundsen, D. C. Ralph, and R. A. Buhrman, Phys. Rev. Lett. **109**, 096602 (2012).

Figures and captions:

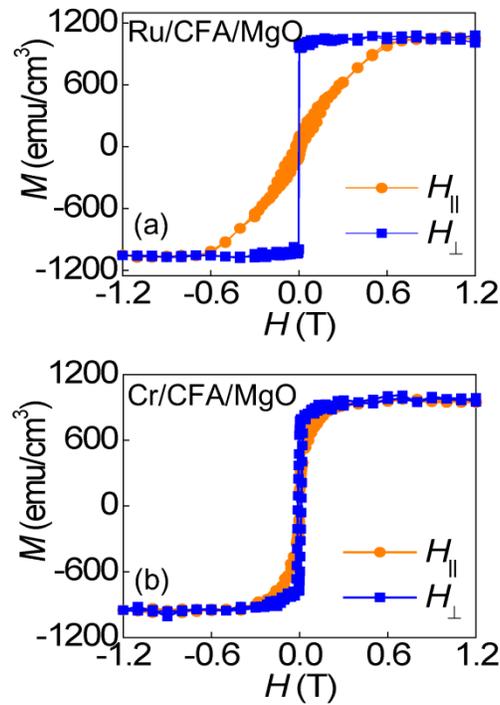

FIG. 1. In-plane and out-of-plane *M-H* loops for CFA films: (a) Ru (0−10)/CFA(1)/MgO(2)/Ta(1) (sample A) and (b) Cr(0−10)/CFA(1)/MgO(2)/Ta(1) (sample B). The thickness is in nanometer.

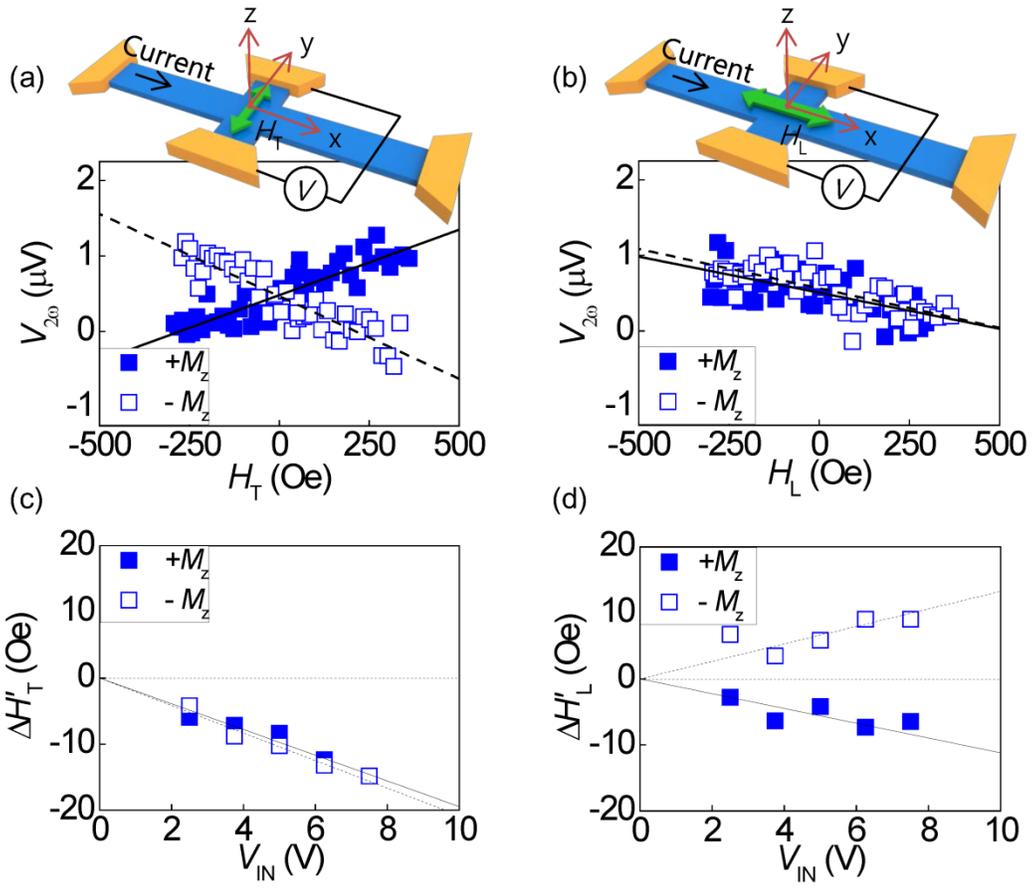

FIG. 2. (a, b) Illustrations of the Hall bar devices and the second harmonic Hall signal $V_{2\omega}$ under external fields directed (a) transverse to and (b) along the current flow measured for sample B (Cr thickness: 8.0 nm, $V_{IN}$ = 7.5 V). (c, d) Transverse and longitudinal effective field $\Delta H'_{T(L)}$ as a function of voltage for sample B (Cr thickness: 8.0 nm). The solid and opened symbols correspond to magnetization orientation along $+M_z$ and $-M_z$.

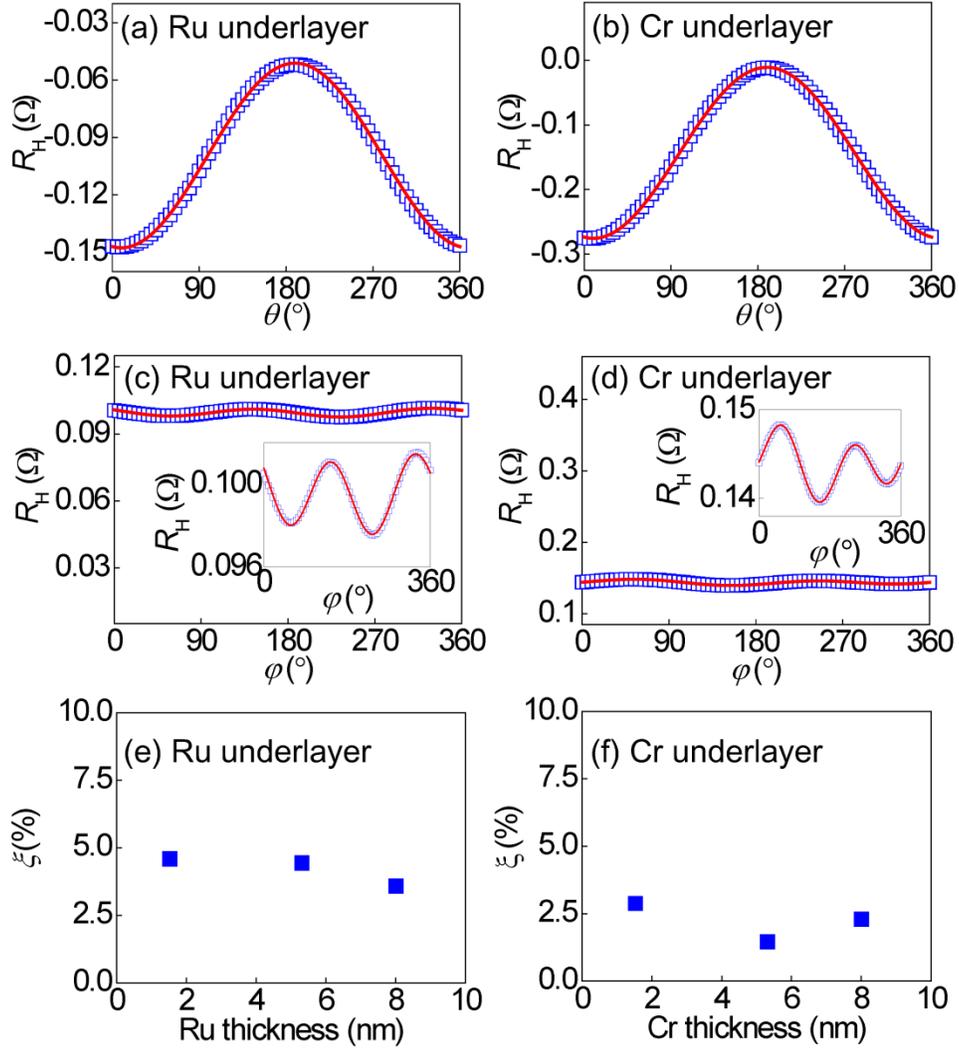

FIG. 3. (a, b) The polar angle $\theta$ dependence of $R_H$ for the (a) sample A and (b) sample B. (c, d) The azimuthal angle $\varphi$ dependence of $R_H$ for the (c) sample A and (d) sample B. Inset is a magnification of the main panel. Results are from devices with 8.0-nm-thick underlayers. The solid lines indicate fitting curves using equation (2). (e, f) The ratio $\xi$ from (e) sample A and (f) sample B as a function of the underlayer thickness.

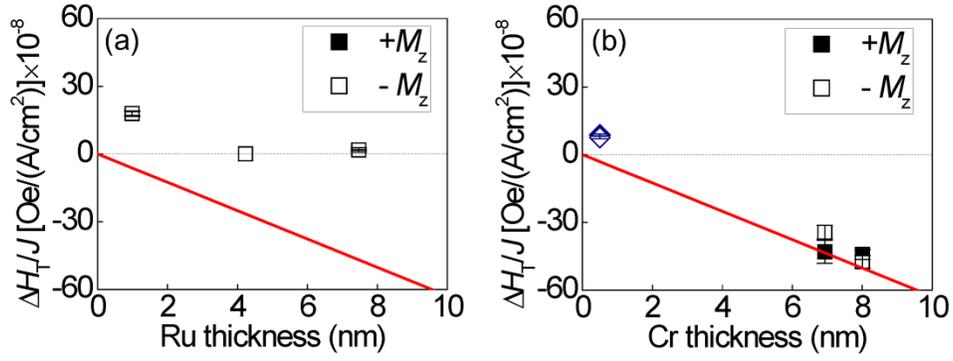

FIG. 4. $\Delta H_T/J$ after the PHE correction in (a) sample A, and (b) sample B and C (diamond symbol) as a function of the underlayer thickness. Solid red lines indicate the estimated Oersted field.

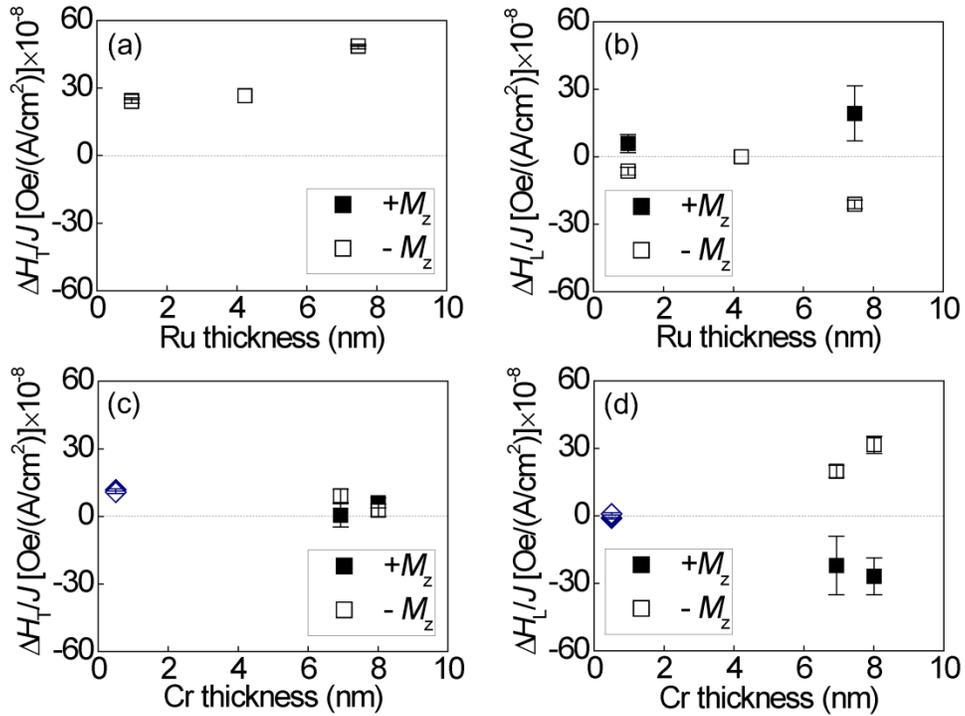

FIG. 5. (a, b) Ru underlayer thickness dependence of (a) $\Delta H_T/J$ and (b) $\Delta H_L/J$ for sample A. (c, d) Cr underlayer thickness dependence of (c) $\Delta H_T/J$ and (d) $\Delta H_L/J$ for sample B and C(diamond symbol). The solid and opened symbols correspond to the data from $+M_z$ and $-M_z$ magnetization states.